\documentstyle[12pt,aasms4]{article} 
 \def\etal{{et~al.}\ }

 \def\vol#1  {{{#1}{\rm,}\ }}

 \def\etal{et al.\ }

 \def\clock{\count0=\time \divide\count0 by 60
      \count1=\count0 \multiply\count1 by -60 \advance\count1 by \time
      \number\count0:\ifnum\count1<10{0\number\count1}\else\number\count1\fi}

\begin{document}
 \title{Evolution of Lyman Break Galaxies Beyond Redshift Four}
 \author{Renyue Cen\altaffilmark{1}}
 \altaffiltext{1} {Princeton University Observatory, Princeton University,
 Princeton, NJ 08544; cen@astro.princeton.edu}

 \begin{abstract}
 The formation rate of luminous galaxies
 seems to be roughly constant
 from $z\sim 2$ to $\sim 4$ 
 from the recent observations of Lyman break galaxies (LBGs)
 (Steidel \etal 1999).
 The abundance of luminous quasars, on the other hand, appears to 
 drop off by a factor of more than twenty from $z\sim 2$ to $z\sim 5$
 (Warren, Hewett, \& Osmer 1994; 
 Schmidt, Schneider, \& Gunn 1995).
 The difference in evolution
 between these two classes of objects
in the overlapping, observed redshift range, $z=2-4$,
 can be explained naturally,
 if we assume that 
 quasar activity is triggered by mergers
 of luminous LBGs and one quasar lifetime is  $\sim 10^{7-8}~$yrs.
 If this merger scenario holds at higher redshift,
 for the evolutions of these two classes of objects
 to be consistent at $z>4$,
 the formation rate of luminous LBGs 
 is expected to drop off at least 
 as rapidly as $\exp\left(-(z-4)^{6/5}\right)$ at $z>4$.

 \end{abstract}


 \section{Introduction}

 Observations of galaxies in the rest frame UV band
 (\cite{Lilly96}1996; \cite{Madau96}1996; 
 \cite{Connolly97}1997;
 \cite{Sawicki97}~1997; \cite{Treyer98}1998; \cite{Pascarelle98}~1998)
 indicate that the galaxy formation rate rises
 steeply from $z=0$ to $z\sim 1$,
 with a nearly constant rate thereafter up to
 $z\sim 4$ 
(\cite{Steidel99}1999).
 While at low redshift ($z<2$) the evolution of luminous
 quasar abundance 
 resembles that of luminous galaxies (e.g., Sanders \& Mirabel 1996;
 Boyle \& Terlevich 1998),
 at high redshift ($z>2$) the two classes of objects do not seem to
 parallel one another, with the
 luminous quasar formation rate (e.g., Warren \etal 1994; Schmidt \etal 1995)
 dropping off more steeply than that
 of luminous galaxies.

 In this {\it Letter}
 a phenomenological approach is taken 
 to relate the {\it observed} formation 
 rate of luminous LBGs
 to the {\it observed} abundance evolution of luminous quasars
 at $z>2$.
 It is shown that, {\it if 1) quasar activity is triggered
 by LBG mergers} and {\it 2) each quasar period lasts
 $\sim 10^{7-8}$yrs},
 then the apparent difference,
 {\it in both shape and amplitude},
 between the evolutions of bright LBGs and bright quasars 
 from $z=2$ to $z=4$ can be explained quantitatively.
 The first assumption finds its support from 
 both the observational evidence
 that a significant fraction of quasar hosts have disturbed
 morphologies or ongoing galaxy-galaxy
 interactions (e.g., Boyce \etal 1996; Bahcall \etal 1997;
 Boyce, Disney, \& Bleaken 1999) 
 and the theoretical consideration that merger of 
 two (spiral) galaxies seems to provide a natural mechanism to fuel 
 the central black hole (e.g., Barnes \& Hernquist 1991).
 The second assumption
 is also theoretically well motivated (Rees 1984, 1990)
 and now strongly implied or required 
 by the mounting observational evidence that most nearby
 massive galaxies seem to harbor inactive black holes at
 their centers (e.g., Richstone \etal 1998).

 The primary purpose of this work is to use this merger
 model to infer the LBG formation rate at higher redshift $z>4$.
 Given the precipitous drop-off of luminous quasar ($M_B<-26.0$)
 abundance from $z=2$ to $z=5$
 the formation rate of luminous ($M_{AB}\geq -23$ to $-22$)
 LBGs
 at higher redshift ($z>4$)
 is predicted to drop off as least as fast as
 $\propto \exp\left(-(z-4)^{6/5}\right)$, 
 if the merger scenario holds.
 A cosmological model with $q_0=0.5$ and 
 Hubble constant $H_0=50$km/sec/Mpc is assumed 
 for the analysis presented here.

 It is noted that this simple merger model
 would probably fail at $z<2$
 without having taken into account 
the evolution of gaseous fuel supply to the central black holes
in galaxies (Kauffmann \& Haehnelt 1999; Haiman \& Menou 1999).

 \section{Galaxy Merger Rate and Quasar Abundance}

Denoting $f(z)$ as the galaxy formation rate (galaxy formation per 
unit time per unit comoving volume) as a function of redshift,
then the (cumulative) number density of formed galaxies 
(number of galaxies per unit comoving volume) is 
(ignoring the small fraction of galaxies that merge)
 \begin{equation}
 g(z)=\int_\infty^z f(z^\prime) {dt\over dz^\prime} dz^\prime.
 \end{equation}
 \noindent 
 For simplicity $\Omega_0=1$ will be assumed, which should be a good
 approximation at high redshift ($z>2$)
 for the range of cosmological models of current interest 
 ($\Omega_0>0.2$).
 The merger rate for a galaxy with an internal
 one-dimensional velocity dispersion $\sigma_i(z)$
 in a cluster/group
 with galaxy number density $d(z)$ (assuming all galaxies under
 consideration are identical)
 and one-dimensional velocity dispersion $\sigma_e(z)$
 has been computed by Makino \& Hut (1997, equation 33) to be:
 \begin{equation}
 P(z)= {18\over \sqrt{\pi}} {1\over x(z)^3} d(z) r_v(z)^2 \sigma_i(z) R(x),
 \end{equation}
 \noindent 
 where $R(x)$ is a dimensionless function of
 $x(z)\equiv \sigma_e/\sigma_i$ which depends on the galaxy model
 and $r_v(z)$ is the virial radius of a galaxy.
 Makino \& Hut (1997) demonstrate that
 $R(x)$ is a constant  ($\sim 11-14$) 
 to good accuracy for $x>2$
 for several different galaxy models.

 Clearly, not all galaxies participate in merging at any given time;
 most galaxies have merger time scales much longer
 than the Hubble time.
 Rather, only galaxies in dense environments 
 such as groups or clusters of galaxies
 have significant probability to merge with others.
 To make the problem more tractable
 it is assumed that 
 a fraction, $\beta(z)$, of 
 all galaxies [$g(z)$] under consideration at any given time
 is in dense environments (i.e., typical groups/clusters at $z$)
 where most mergers occur,
 and the remainder of galaxies (i.e., field galaxies)
 have zero probability of merger.
 Then, the total merger rate is
 \begin{equation}
 M(z) = \beta(z) g(z) P(z)
 \end{equation}
 \noindent 
 and the quasar abundance at any given redshift $z$ is

 \begin{equation}
 Q(z) = M(z) t_Q(z)
 \end{equation}
 \noindent 
 where $t_Q(z)$ is the assumed quasar lifetime
 (assuming that $t_Q$ is much less than the Hubble time, which
 turns out to be necessary for the model to be viable).

 There are two significantly uncertain
 remaining parameters, $d(z)$ and $\sigma_e(z)$,
 which need to be specified.
 It is noted that 
 quasar activities at high redshift seem to occur mostly in regions
 with galaxy number density typical of present-day
 clusters/groups of galaxies.
 This information is provided by observations of quasar companions
 which have a typical separation from a quasar of a few hundred 
 comoving kiloparsecs
 (e.g., Djorgovski 1999).
 At redshift $z=1-2$
 there is evidence from larger observational data sets
 that quasars reside
 in cluster-like environment (Hall \& Green 1998).
 It thus appears that $d(z)$
 may be a weak function of redshift and
 is assumed to be constant here (more discussion on this later).
 The velocity dispersion of characteristic
 systems (groups/clusters in this case),
 $\sigma_e(z)$, should be a decreasing
 function of redshift in any hierarchical cosmological model.
 Here we take advantage of the insight of Kaiser (1986)
 and use the solution for simple power-law
 models:
 \begin{equation}
 \sigma_e(z) = \sigma_e(0) (1+z)^{{1\over 2} {n-1\over n+3}},
 \end{equation}
 \noindent 
 where $n$ is the power index of the primordial
 density fluctuation spectrum at the relevant scales for clusters/groups.
 For cold dark matter like models or from observations of local
 large scale structure $n$ is expected to be $\sim -1$.

 The purpose of estimating LBG formation rate
 at $z>4$ 
 is met by finding $f(z)$ at $z>4$ that matches the observed
 quasar abundance in the range $z>2$.
For the present analysis a simple functional
form of LBG formation rate is adopted:
\begin{eqnarray}
&f(z)&=A\quad \quad\quad \quad \quad \quad \quad \quad \quad \quad \quad \quad\hbox{for} \quad \quad \quad 2<z<4\nonumber\\
&f(z)&=A\exp \left(-(z-4)^{6/5}\right) \quad\quad \quad \quad\hbox{for} \quad \quad \quad z>4,
\end{eqnarray}
\noindent 
consistent with the latest LBG observations at high redshift up
to $z=4$ (Steidel \etal 1999),
where $A$ is a normalization constant.
 At $z>4$, where observations are unavailable,
 a simple form is proposed so as to provide
 an adequate fit to the observed quasar abundance at $z>4$ (see Figure 1 below).
 Using equations (1-3,5-6),
 we find $Q(z)$ (equation 4),
 shown as the heavy solid curve in Figure 1.
 Here, for the shown $Q(z)$ we use
 $n=-1.0$,
 $\sigma_e(0)=10^3$km/s,
 $\sigma_i=100$km/s,
 $\beta=0.025$ (being constant which is consistent with the adoption
 of $n=-1$ powerlaw model),
 $d=40.0~h^{3}$Mpc$^{-3}$,
 $r_v=200h^{-1}$kpc,
 $R(x)=12$ 
 and $t_{Q}=3\times 10^{7}$yrs.
 A cosmological model with $q_0=0.5$ and 
 Hubble constant $H_0=50$km/sec/Mpc is assumed.
 Also shown as symbols are observational data
 of bright quasars ($M_B<-26.0$):
 open circles are from Warren \etal (1994)
 and
 solid dots are from Schmidt \etal (1995).
The open square from Kennefick, Djordovski, \& de Carvalho (1995)
for $M_B<-26.7$ quasars is shown to indicate
the steepness of quasar luminosity function
near the absolute magnitude $M_B\sim -26.0$.

 It is seen that the merger model provides an adequate fit to the
 observed luminous
 quasar abundance in the entire redshift range considered ($z>2$).
 The dashed curve in Figure 1 shows $f(z)$ with
 arbitrary vertical units.
 The dotted curve in Figure 1 shows $g(z)$,
 normalized to be $1.0\times 10^{-4}h^3$Mpc$^{-3}$
 at $z\sim 3$. 
 Note that Figure 5 of Steidel \etal (1999)
 shows the differential  luminosity function 
 of UV bright LBG galaxies (i.e., star-forming galaxies), 
 calling it $g_d(z)$,
 while here, $g(z)$ is the cumulative density of formed galaxies. 
 Roughly speaking, if $f(z)$ is constant,
 then $g(z)/g_d(z) = t_H(z)/t_{SF}$,
 where $t_H(z)$ is the Hubble time at redshift $z$
 and $t_{SF}$ is the star (burst) formation duration 
 (i.e., LBG phase) of a galaxy.
 Since $t_H(z)/t_{SF}\approx 10^9 yrs/10^8 yrs \approx 10$,
the above normalization 
 roughly corresponds to LBGs with $g_d(z)\sim 10^{-5}h^3$Mpc$^{-3}$,
 which in turn corresponds to LBGs with $M_{AB}=-23$ to $-22$
 (Figure 5 of Steidel \etal 1999).

 \section{Discussion}

 On one hand, as $Q(z)$ at $z<4$ does not depend sensitively on
 the form of $f(z)$ at $z>4$,
 the good agreement
 between $Q(z)$ and the observed quasar abundance 
 in the redshift range $z=2-4$ 
 (where both types of objects are observed)
 suggests that the merger scenario 
 of luminous LBGs provides a quantitatively viable model
 for bright quasar formation.
 On the other hand, $Q(z)$ at $z>4$ does depend sensitively on
 the adopted form of $f(z)$ at $z>4$. 
The fact that the proposed model yields
an overall shape at $z=2-5$ that fits observations
implies that the luminous LBG formation rate should
drop off at $z>4$ as indicated by $f(z)$ in eq. 1,
{\it if merger scenario holds at $z>4$}.
But to have a secure estimate of $f(z)$ at $z>4$,
it is vital to understand
the dependences of $Q(z)$ on various other parameters,
namely,
$Q(z)\propto \beta(z) d(z) \sigma_i^4(z) r_v^2(z) t_{Q}(z)(1+z)^{-{3\over 2}{n-1\over n+3}}$. 
We have set each of the parameters constant
(independent of redshift),
which is considered to {\it conservative} in the following discussions
{\it if} a more likely redshift dependence of the quoted parameter
(holding all other parameters constant)
would require an even steeper decreasing function for $f(z)$ at $z>4$
than indicated by equation (6).
Let us now examine each parameter to assess 
how each parameter may vary with redshift.

First, it seems that $\sigma_i(z)$, $r_v(z)$ and $t_{Q}$
are likely to decrease with redshift, making the assumption of their
being constant {\it conservative}.

Second, $\beta=0.025$ is equivalent to 
the assumption of mergers taking place in galaxy systems
corresponding roughly to $2\sigma$ peaks and has 
implications for the correlation function of quasars.
The bias factor of halos 
over mass is $b=1+(\nu^2-1)/\delta_c$ (Mo \& White 1996),
equal to $2.91$ for $\nu=2$ and $\delta_c=1.57$.
If the cluster-cluster correlation function
has a shape $\propto r^{-2}$ (close to the usual
slope of $-1.8$), then the correlation
length of clusters is $b r_{m}$, where $r_{m}$ is the
correlation length of the underlying mass and
evolves as $\propto (1+z)^{-1}$ (Kaiser 1986)
for $n=-1$ and $\Omega_0=1$. 
Our choice of $\beta=0.025$ consequently
implies a correlation length for quasars of approximately
$2.91 r_{m}(0) /(1+z)$, which is equal to $\sim 5h^{-1}$ comoving Mpc
at $z\sim 2$ (using $r_{m}(0)\sim 5.0h^{-1}$Mpc),
in agreement with what is observed for quasars
(e.g., Kundic 1997; Boyle \etal 1998).
In any case, it is unlikely that $\beta$ decreases with redshift.
Therefore, setting $\beta(z)$ constant is {\it conservative}.
An important implication of this model 
is that the comoving correlation length
of luminous quasars should {\it decrease} with redshift
no faster than $(1+z)^{-1}$ at $z>2$,
a potentially testable prediction.
Stephens \etal (1997) give a correlation length 
of $z>2.7$ quasars of $17.5\pm 7.5h^{-1}$Mpc.
It will be very valuable to determine the correlation
length of high redshift quasars with significantly smaller errorbars.

Third, observations may have indicated
that $d(z)$ may be an increasing function
of redshift at $z>4$ (Djorgovski \etal 1997; Djorgovski 1999).
Therefore, assuming $d(z)$ to be constant is {\it conservative}.

Finally, for a plausible power spectrum (such as CDM like)
$n$ is likely to be smaller at smaller scales
thus smaller at higher redshift.
Thus, assuming $n$ to be a constant is {\it conservative}.
Overall, our assumption of constancy for various parameters 
seems {\it conservative}; i.e., $f(z)$ should decrease
at least as rapidly as indicated by equation (6) at $z>4$.




All the analyses so far have been based on the 
available (optical) observations of quasars, which
appears to indicate a sharp drop-off of quasar abundance at $z>4$.
Dust obscuration effects
are often invoked to explain the apparent drop-off of quasar
abundances at high redshift (e.g., Ostriker \& Heiler 1984; Pei 1995).
However, recent radio surveys of high redshift quasars
seem to indicate that the drop-off of the number density
of bright radio quasars is very similar to that
from optical surveys (e.g., Hook, Shaver, \& McMahon 1998)
with the implication that the effect of dust on
the observed drop-off of bright quasars at $z>2$
may be small.

One potential problem with the merger model is
that observations show that a large fraction of 
quasar hosts at low redshift ($z<0.5$)
appear to be quite normal looking, i.e., without disturbed appearances.
But one would expect that, if galaxy-galaxy merger time scale
is longer than the proposed quasar lifetime,
all quasar hosts should display appearances of some interaction.
One possible solution to this problem is that
quasar formation is delayed, i.e., a quasar does not start to shine
until the galaxy-galaxy merger is nearly complete.
In other words, the time it takes to set up 
the central (BH) region for quasar activity  during galaxy merger
may be comparable to the time that it takes
for the two galaxies to merger.

\section{Conclusions}

In an early classic paper Efstathiou \& Rees (1988) show
that quasar abundance at high redshift can be accounted for
in the standard cold dark matter model {\it if massive halos
are related to the formation of black holes},
with an intriguing prediction 
that the abundance of luminous quasars should decrease rapidly beyond $z=5$
(for a more recent treatment see Haehnelt \& Rees 1993).
[The evolution of low-luminosity quasars, of course, 
does not necessarily have to follow
that of their luminous counterparts (e.g., Haiman \& Loeb 1998)].

 In this {\it Letter} a different approach is taken by
 directly relating the {\it observed} evolution of luminous LBGs
 to the {\it observed} evolution of luminous quasars
 at high redshift ($z>2$).
 With a set of seemingly reasonable parameter values,
 it is shown that consistency between the two classes of objects
 at $z=2-4$, where both classes are observed,
 can be achieved, if one assumes that 1) {\it Lyman break galaxies
 merger to trigger quasar activity} and 2) {\it quasar lifetime
 is $\sim 10^{7-8}$yrs}. 
 At $z>4$, consistency
 can be achieved, {\it only if additionally}
 the formation rate of luminous LBGs drops off
 as $\exp(-(z-4)^{6/5})$ or faster,
 a prediction that may be tested by future observations.
 One implication from this model 
 is that LBGs with $M_{AB}\geq -23$ to $-22$
 merge to form quasars with  $M_B<-26.0$ at $z>2$.
 Correlation analysis of relevant LBGs and quasars should
 shed light on this.

 At lower redshift additional, more model dependent
 assumptions regarding the supply of available gas
 to fuel black holes would be required
 to make qualitatively viable predictions.
 Kauffmann \& Haehnelt (1999; see also Haiman \& Menou 1999)
 have presented
 a detailed model, based also on merger scenario,
 to unify the evolution of galaxies and quasars
 in the cold dark matter model under several plausible assumptions
concerning the evolution of fuel gas to the central black holes.
The success of the model of 
 Kauffmann \& Haehnelt (1999)
 at low redshift ($z<2$)
 and the model presented here at high redshift
 ($z>2$), both based on galaxy merger scenario,
suggests that {\it galaxy merger may play an indispensable
role in quasar formation.}

 \acknowledgments
 The work is supported in part
 by grants AST9318185 and ASC9740300.
 I thank Xiaohui Fan, Zoltan Haiman, Jerry Ostriker,
 Michael Strauss and David Weinberg 
 for many useful discussions.
 An anonymous referee is acknowledged
 for helpful comments.

 {}
 \newpage
 \figcaption[FLENAME]{
 The heavy solid curve shows the evolution
 of the abundance of bright quasars computed using the
 merger model of this paper.
 Also shown as symbols are observational data
 of bright quasars ($M_B<-26.0$):
 open circles are from Warren \etal (1994)
 and
 solid dots are from Schmidt \etal (1995).
The open square from Kennefick, Djordovski, \& de Carvalho (1995)
for $M_B<-26.7$ quasars is shown to indicate
the steepness of quasar luminosity function
near the absolute magnitude $M_B\sim -26.0$.
 The dotted and dashed curves show $g(z)$ (eq. 1)
 and $f(z)$ (eq. 6), respectively.
 \label{fig1}}

 
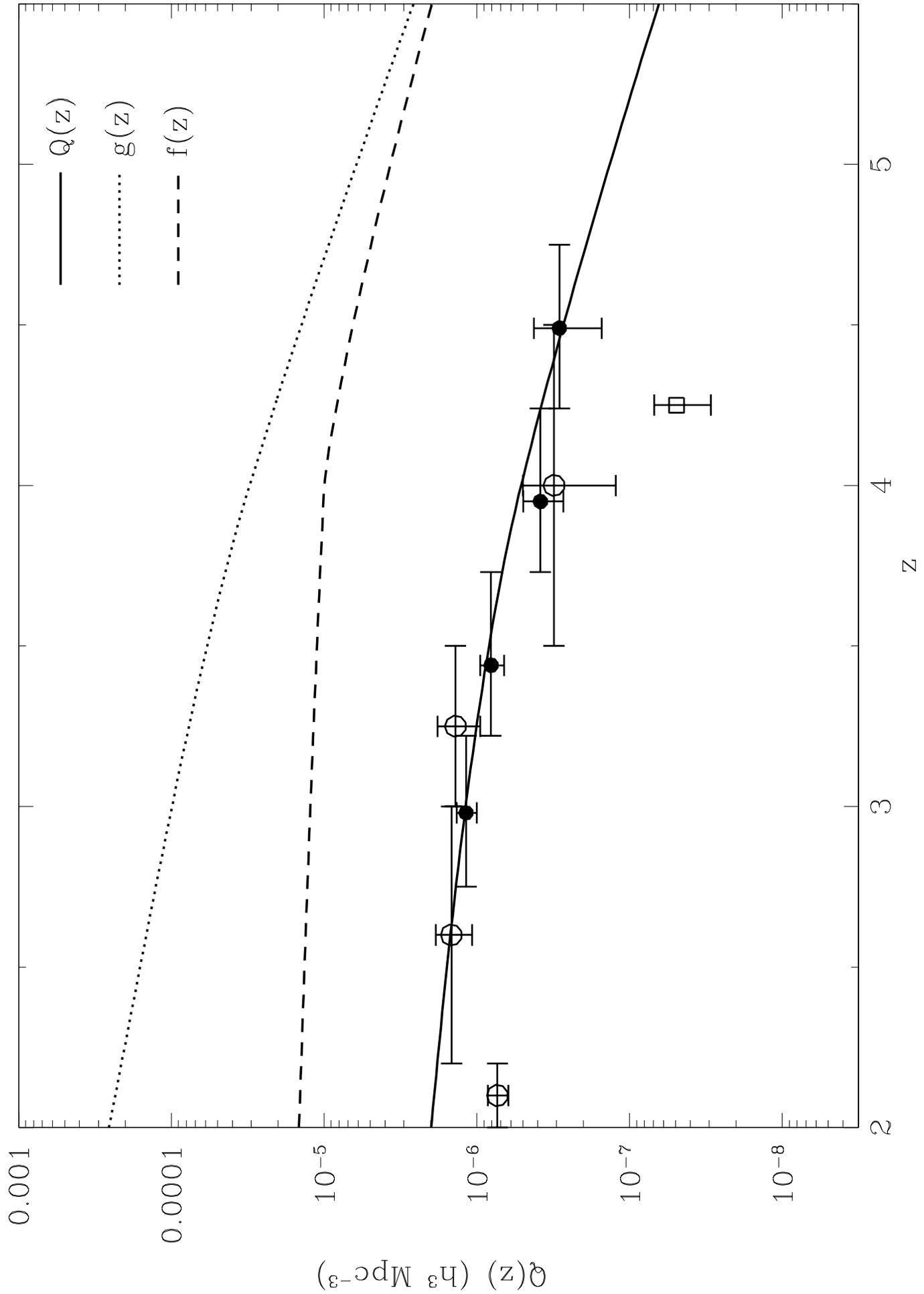
\end{document}